# Testing of MOND with Local Group Timing

Yan-Chi Shi
yanchishi@yahoo.com

**Abstract**

The timing of the local group is used to test Modified Newtonian Dynamics (MOND). The result shows that the masses predicted by MOND are well below the baryonic contents of the Milky Way and the Andromeda galaxies.

**Key Words:** dark matter – Local Group – galaxies: interactions – galaxies: kinematics and dynamics

## 1. Introduction

Although dark matter has not been directly detected yet, there is strong evidence for its existence in the universe (Zwicky 1933, Rubin & Ford 1972). Recent reports showed that dark matter is essential in the formation of the large-scale structure of the universe (Springel, Frenk & White 2005).

At the largest scale, the flatness of the observed universe and the cosmic microwave background (CMB) data requires 23% of the mass-energy density in the universe to be dark matter (Komatsu et al 2008). However, attempts to directly detect dark matter have failed despite decades of effort.

It is reasonable to raise a question about dark matter being another kind of "ether" which was widely accepted at the end of 19th century. The need for dark matter is based on Newton's law of gravitation and theory of general relativity, which are well-tested in the lab and up to the scale of the solar system. There is a possibility that both theories cannot be extended in some conditions. The most successful alternative to dark matter is Modified Newtonian Dynamics (MOND) as proposed by Milgrom (1983).

MOND is extremely successful in explaining the discrepancy between the rotation and luminosity curves in spiral galaxies. In other words, MOND requires only the baryonic masses to explain the flat rotation curves of spiral galaxies. There is no need for the existence of dark matter in the galaxy scale (Sanders & McGaugh 2002). Numerous observational data were used to test MOND in the last two decades (Pointecouteau & Silk 2005, Klypin & Prada 2009). Very few of them provide conclusive evidence to falsify MOND. The most convincing one is the recent gravitational lensing study of 1E 0657-558 also known as the Bullet Cluster (Clowe et al 2006). The study shows that the location containing the most baryonic matter does not match the gravitational lensing centers, where most of the mass lies, after the collision of two galaxy clusters. This is viewed as convincing evidence for the existence of dark matter.

In this paper I apply the timing of the local group to test MOND. I find that MOND predicts a much smaller mass for the Milky Way. The predicted mass is a factor of four smaller than the baryonic mass of the Milky Way.

In Section 2, I repeat the calculation using the Newtonian motion and thus determine the timing between Andromeda (M31) and the Milky Way. I obtain the same results as in many previous papers (Kahn & Woltjer 1959; Mishra 1985; Binney & Tremaine 2008).

In Section 3, MOND is used to calculate the motion and timing of these two galaxies.

## 2. Newtonian

The current separation between M31 and the Milky Way is about 740 Kpc (Binney & Tremaine 2008; Ribas et al 2005). They are moving toward each other with a speed of 125 Km $s^{-1}$ (Binney & Tremaine 2008). According to the Big Bang theory, these two galaxies started out close to each other, and then moved apart due to the expansion of the universe. The gravitational attraction slowed them down, halted, and then reversed their recessional motion. If we treat the Milky Way (with mass $m_1$) and M31 (with mass $m_2$) as two point masses and assume they are moving in an almost radial orbit, their motion is determined by Newton's law of gravitation. If we choose a coordinate system where the origin is at the center of mass of the Milky Way and M31 system, and $r_1$ is the distance of the Milky Way to the origin, then

$$m_1\frac{d^2r_1}{dt^2} = -G\frac{m_1m_2}{r^2}.$$

Similarly,

$$m_2\frac{d^2r_2}{dt^2} = -G\frac{m_1m_2}{r^2}, \quad \text{or}$$

$$\frac{d^2r}{dt^2} = -G\frac{M}{r^2} \qquad \text{where } r = r_1 + r_2,\ M = m_1 + m_2 \qquad (1)$$

At $r_m$ (the maximum separation between M31 and the Milky Way), $\dot{r} = 0$.

$$\dot{r}^2 = \frac{2GM}{r} - \frac{2GM}{r_m} \qquad (2)$$

$$\int \frac{\sqrt{r}\,dr}{\sqrt{r_m - r}} = -\sqrt{\frac{2GM}{r_m}}\int dt$$

By integration of the above equation we can find the time $t_{0m}$ for M31 and the Milky Way to move from $r = 0$ to $r_m$.

$$t_{0m} = \frac{\pi\, r_m^{3/2}}{2\sqrt{2GM}}$$

Likewise, we can find the time $t_{mn}$ for M31 and the Milky Way to move from $r_m$ to the separation now $r_n$ ( =740 Kpc).

The total time for M31 and the Milky Way to move from to $r = 0$ to $r_m$ and from $r_m$ to the current separation $r_n$

$t_{total} = t_{0m} + t_{mn}$ = 13.7 Gyr = the age of the universe

$$= \frac{\pi\, r_m^{3/2}}{\sqrt{2GM}} + \sqrt{\frac{r_m}{2GM}}\left[\sqrt{r_n}\sqrt{r_m - r_n} - r_m \tan^{-1}\frac{\sqrt{r_n}}{\sqrt{r_m - r_n}}\right] \qquad (3)$$

From Equation 2, we can get $r_m$

$$\dot{r}_n^{\,2} = \frac{2GM}{r_n} - \frac{2GM}{r_m} \qquad (4)$$

where $r_n$ = 740 Kpc is the separation between M31 and the Milky Way at the present time, and $\dot{r}_n$ = 125 Km $s^{-1}$ is the corresponding speed.

From Equations 3 & 4, we can find M numerically.

$M \equiv m_1 + m_2 = 4.5 \times 10^{12}\ M_\odot$

This is far larger than the baryonic mass in these two galaxies which is in the order of $10^{11}\ M_\odot$ (Binney & Tremaine 2008). This is considered as a strong evidence of the existence of dark matter.

## 3. Modified Newtonian Dynamics (MOND)

The original goal of MOND is to use only the observed matter to explain the flat rotation curves of spiral galaxies and the well-known Tully-Fisher relation. It is phenomenological rather than based on any sound physical hypothesis or principle. It formulates mathematical equations to fit the flat rotation curves of spiral galaxies. Regardless of this, MOND is quite successful and works well when applied to galaxies (Sanders 2008).

If $\mathbf{g_m}$ is the MOND gravitational acceleration, and $\mathbf{g_n}$ is the Newtonian acceleration, Milgrom (Milgrom 1983; Sanders & McGaugh 2002) proposed that

$$\mathbf{g_m}\ \mu ( | \mathbf{g_m} | / a_0 ) = \mathbf{g_n} \ , \qquad (5)$$

where $\mu (x) = 1$, when $x >> 1$ (the Newtonian limit),

$\mu (x) = x$, when $x << 1$ (deep-MOND regime),

$a_0 = 1.2 \times 10^{-8}$ cm s $^{-2}$ is the MOND parameter which has the unit of acceleration (Sanders & McGaugh 2002).

It was recognized that Equation 5 does not preserve the conservation of linear momentum (Felten 1984). In 1984 Bekenstein and Milgrom proposed a rigorous Lagrangian formulation of MOND. The modified Poisson equation becomes

$$\nabla \cdot [ \mu ( | \nabla \phi | / a_0 ) \nabla \phi ] = 4\pi G\rho, \qquad (6)$$

where $-\nabla \phi = \mathbf{g_m}$.

This modification conserves linear momentum and energy as it is derived from Lagrangian formulation.

In deep-MOND regime, for high symmetry cases where the density distribution is spherical, cylindrical, or planar, Equation 6 reduces to Equation 5 which provides a much simpler algebraic relation between the MOND acceleration and the Newtonian acceleration.

Thus, in the limit of low accelerations (deep-MOND regime), from Equation 5 the MOND acceleration is given by

$$g_m = \sqrt{a_0 g_n}$$

The above simplified equation is a good approximation to compute the timing between M31 and the Milky Way, since most of the time the system is in the deep-MOND regime except for a very short period of time when these two galaxies were close to each other.

The MOND equation of the motion between M31 and the Milky Way is given by

$$\frac{d^2r}{dt^2} = -\sqrt{a_0 g_n}\ ,$$

where $g_n$ is the Newtonian acceleration given by Equation 1, and $r$ is the separation between M31 and the Milky Way.

$$\frac{d^2r}{dt^2} = -\frac{\sqrt{a_0 G}}{r}\left(\sqrt{m_1 + m_2}\ \right)$$

This equation can be written as

$$\frac{d^2r}{dt^2} = -\frac{\sqrt{a_0 G m_1}}{r}\eta\ , \quad \text{where} \quad \eta \equiv \sqrt{1 + \frac{m_2}{m_1}}$$

At $r_m$ (the maximum separation between M31 and the Milky Way), $\dot{r} = 0$.

$$\dot{r}^2 = 2\,\eta\,\sqrt{a_0 G m_1}\ (\ln r_m - \ln r) \qquad (7)$$

$$\int \frac{dr}{\sqrt{\ln r_m - \ln r}} = -\sqrt{2\eta}(Gm_1 a_0)^{1/4} \int dt$$

The left hand side $= -r_m\sqrt{\pi}\ \mathrm{erf}\left(\sqrt{\ln r_m - \ln r}\ \right)$.

The time for M31 and the Milky Way to move from $r = 0$ *to* $r_m$ (the maximum separation)

$$t_{0m} = \frac{r_m\sqrt{\pi}}{\sqrt{2\eta}(Gm_1a_0)^{1/4}}$$

The time for M31 and the Milky Way to move from $r_m$ to the current separation $r_n$

$$t_{mn} = \frac{r_m\sqrt{\pi}}{\sqrt{2\eta}(Gm_1a_0)^{1/4}} \text{ erf}\left(\sqrt{\ln r_m - \ln r_n}\right)$$

The total time for M31 and the Milky Way to move from $r = 0$ to $r_m$ and from $r_m$ to the current separation $r_n$

$t_{total} = t_{0m} + t_{mn} = 13.7$ Gyr

$$= \frac{r_m\sqrt{\pi}}{\sqrt{2\eta}(Gm_1a_0)^{1/4}} \left[1 + erf\left(\sqrt{ln\, r_m - ln\, r_n}\right)\right] \qquad (8)$$

The maximum separation $r_m$ could be obtained by inserting the current separation $r_n = 740$ Kpc and the speed $\dot{r}_n = 125$ Km s$^{-1}$ into Equation 7.

$$\dot{r}_n{}^2 = 2\,\eta\,\sqrt{a_0Gm_1}\,(\ln r_m - \ln r_n) \qquad (9)$$

From Equations 8 & 9, we can find the MOND mass $m_1$ of the Milky Way numerically.

In order to find $m_1$ we need to provide a value for

$$\eta \equiv \sqrt{1 + \frac{m_2}{m_1}}.$$

Assuming the mass of M31 ($m_2$) is about 1.5 times the mass of the Milky Way ($m_1$), we get

$m_1 = 1.2 \times 10^{10}\ M_\odot$ .

This is only 24% of the estimated baryonic mass of the Milky Way which is about $5 \times 10^{10}\ M_\odot$ (Binney & Tremaine 2008) .

Instead of using Equation 8 to calculate the masses, Equation 8 can be used to calculate the time required for M31 and the Milky Way to get to the current separation if the masses for the Milky Way and M31 are known. If only the baryonic masses of the Milky Way (5 X $10^{10}$ $M_\odot$) and M31 (1.5 X 5 X$10^{10}$ $M_\odot$) are used, the time required for M31 and the Milky Way to reach to their current separation is 7.3 Gyr. Thus MOND's prediction of the age of the universe is much shorter than the believed age of the universe. This is understandable because the MOND force is stronger than the Newtonian force in the deep-MOND regime. It would take less time to complete the same motion.

## 4. Discussion

One of the uncertainties of the above argument is that we do not know the exact past history of the local group. For example, if one assumes that M31 and the Milky Way had already completed one orbit and are approaching the completion of the second orbit now, the MOND calculation shows that the total time is 17.5 Gyr. This is not consistent with the current age of the universe either.

The other uncertainty is the assumption that M31 and the Milky Way are moving in an almost radial orbit. If the orbit is non-radial, the timing will be longer. In order to increase the timing from 7.3 Gyr (as predicted by MOND for the radial orbit) to 13.7 Gyrs, there should be a considerable amount of proper motion. The required proper motion will be in the same order of magnitude as the current radial velocity between M31 and the Galaxy. That is the proper motion in the order of $10^2$ Km $s^{-1}$. This is inconsistent with the current view of the universe that the initial motion between M31 and the Milky Way is due to Hubble expansion in the early universe which has zero angular momentum. There are only two massive members, M31 and the Milky Way, in the Local Group. It is hard to explain how the system acquired so much angular momentum.

It is interesting to point out that the proper motion of $10^2$ Km $s^{-1}$ is equivalent to $10^{-5}$ arcseconds position change per year of M31in the sky. This will be in the measurable range of SIM (Space Interferometry Mission) in the future.

## 5. Conclusion

MOND is extremely successful in explaining the flat rotation curves of spiral galaxies and the Tully-Fisher relation (Sanders & McGaugh 2002) without the need for dark matter. However, when it is applied to galaxy clusters, MOND predicts masses that are double (Sanders 2003) or even several times greater than the baryonic masses (Pointecouteau & Silk 2005). Sanders (2003) argued that MOND's prediction of more mass does not constitute the falsification of MOND because there might be other forms of matter in the system that are not visible (such as neutrinos). His argument is that if MOND predicts less mass than is observed, then it will be a definite falsification of MOND. More mass can always be found, but it is difficult to make observed mass disappear.

In this paper I apply MOND to a simpler case, the motion and timing of M31 and the Milky Way. The result is that MOND predicts much less mass than the known baryonic mass for the Milky Way. The fact that MOND predicts less mass provides a potential problem for MOND.

The manuscript originated from a term paper study of the "Galactic Dynamics" course at Rutgers University. I would like to thank Professor Sellwood for his invaluable comments as well as inspiration. I would also like to thank Rutgers University for letting me audit several astronomy courses in the past few years. I am grateful to Professor Stacy McGaugh for the interesting discussions on non-radial motion.